
%
\magnification=1200
%
%
\hsize=31pc
\vsize=55 truepc
\hfuzz=2pt
\vfuzz=4pt
\pretolerance=5000
\tolerance=5000
\parskip=0pt plus 1pt
\parindent=16pt
%
%
\font\fourteenrm=cmr10 scaled \magstep2
\font\fourteeni=cmmi10 scaled \magstep2
\font\fourteenbf=cmbx10 scaled \magstep2
\font\fourteenit=cmti10 scaled \magstep2
\font\fourteensy=cmsy10 scaled \magstep2
\font\large=cmbx10 scaled \magstep1
%
\font\sans=cmssbx10
%
%
\def\bss#1{\hbox{\sans #1}}
%
%

%
%

%
%
\font\eightrm=cmr8
\font\eighti=cmmi8
\font\eightbf=cmbx8
\font\eightit=cmti8

\font\eightsy=cmsy8
\font\sixrm=cmr6
\font\sixi=cmmi6
\font\sixsy=cmsy6

\def\tenpoint{\def\rm{\fam0\tenrm}%
  \textfont0=\tenrm \scriptfont0=\sevenrm
                      \scriptscriptfont0=\fiverm
  \textfont1=\teni  \scriptfont1=\seveni
                      \scriptscriptfont1=\fivei
  \textfont2=\tensy \scriptfont2=\sevensy
                      \scriptscriptfont2=\fivesy
  \textfont3=\tenex   \scriptfont3=\tenex
                      \scriptscriptfont3=\tenex
  \textfont\itfam=\tenit  \def\it{\fam\itfam\tenit}%
  \textfont\slfam=\tensl  \def\sl{\fam\slfam\tensl}%
  \textfont\bffam=\tenbf  \scriptfont\bffam=\sevenbf
                            \scriptscriptfont\bffam=\fivebf
                            \def\bf{\fam\bffam\tenbf}%
  \normalbaselineskip=20 truept
  \setbox\strutbox=\hbox{\vrule height14pt depth6pt
width0pt}%
  \let\sc=\eightrm \normalbaselines\rm}
\def\eightpoint{\def\rm{\fam0\eightrm}%
  \textfont0=\eightrm \scriptfont0=\sixrm
                      \scriptscriptfont0=\fiverm
  \textfont1=\eighti  \scriptfont1=\sixi
                      \scriptscriptfont1=\fivei
  \textfont2=\eightsy \scriptfont2=\sixsy
                      \scriptscriptfont2=\fivesy
  \textfont3=\tenex   \scriptfont3=\tenex
                      \scriptscriptfont3=\tenex
  \textfont\itfam=\eightit  \def\it{\fam\itfam\eightit}%
  \textfont\bffam=\eightbf  \def\bf{\fam\bffam\eightbf}%
  \normalbaselineskip=16 truept
  \setbox\strutbox=\hbox{\vrule height11pt depth5pt width0pt}}
\def\fourteenpoint{\def\rm{\fam0\fourteenrm}%
  \textfont0=\fourteenrm \scriptfont0=\tenrm
                      \scriptscriptfont0=\eightrm
  \textfont1=\fourteeni  \scriptfont1=\teni
                      \scriptscriptfont1=\eighti
  \textfont2=\fourteensy \scriptfont2=\tensy
                      \scriptscriptfont2=\eightsy
  \textfont3=\tenex   \scriptfont3=\tenex
                      \scriptscriptfont3=\tenex
  \textfont\itfam=\fourteenit  \def\it{\fam\itfam\fourteenit}%
  \textfont\bffam=\fourteenbf  \scriptfont\bffam=\tenbf
                             \scriptscriptfont\bffam=\eightbf
                             \def\bf{\fam\bffam\fourteenbf}%
  \normalbaselineskip=24 truept
  \setbox\strutbox=\hbox{\vrule height17pt depth7pt width0pt}%
  \let\sc=\tenrm \normalbaselines\rm}

\def\today{\number\day\ \ifcase\month\or
  January\or February\or March\or April\or May\or June\or
  July\or August\or September\or October\or November\or
December\fi
  \space \number\year}
%
%
\newcount\secno      
\newcount\subno      
\newcount\subsubno   
\newcount\appno      
\newcount\tableno    
\newcount\figureno   
\normalbaselineskip=20 truept
\baselineskip=20 truept
%
%
\def\title#1
   {\vglue1truein
   {\baselineskip=24 truept
    \pretolerance=10000
    \raggedright
    \noindent \fourteenpoint\bf #1\par}
    \vskip1truein minus36pt}
%
%
\def\author#1
  {{\pretolerance=10000
    \raggedright
    \noindent {\large #1}\par}}
%
%
\def\address#1
   {\bigskip
    \noindent \rm #1\par}
%
%
\def\shorttitle#1
   {\vfill
    \noindent \rm Short title: {\sl #1}\par
    \medskip}
%
%
\def\pacs#1
   {\noindent \rm PACS number(s): #1\par
    \medskip}
%
%
\def\jnl#1
   {\noindent \rm Submitted to: {\sl #1}\par
    \medskip}
%
%
\def\date
   {\noindent Date: \today\par
    \medskip}
%
%
\def\beginabstract
   {\vfill\eject
    \noindent {\bf Abstract. }\rm}
%
%
\def\keyword#1
   {\bigskip
    \noindent {\bf Keyword abstract: }\rm#1}
%
%
\def\endabstract
   {\par
    \vfill\eject}
%
%
%

%
%
\def\entry#1#2#3
   {\noindent
    \hangindent=20pt
    \hangafter=1
    \hbox to20pt{#1 \hss}#2\hfill #3\par}
%
%
\def\subentry#1#2#3
   {\noindent
    \hangindent=40pt
    \hangafter=1
    \hskip20pt\hbox to20pt{#1 \hss}#2\hfill #3\par}
%
%
\def\section#1
   {\vskip0pt plus.1\vsize\penalty-250
    \vskip0pt plus-.1\vsize\vskip24pt plus12pt minus6pt
    \subno=0 \subsubno=0
    \global\advance\secno by 1
    \noindent {\bf \the\secno. #1\par}
    \bigskip
    \noindent}
%
%
\def\subsection#1
   {\vskip-\lastskip
    \vskip24pt plus12pt minus6pt
    \bigbreak
    \global\advance\subno by 1
    \subsubno=0
    \noindent {\sl \the\secno.\the\subno. #1\par}
    \nobreak
    \medskip
    \noindent}
%
%
\def\subsubsection#1
   {\vskip-\lastskip
    \vskip20pt plus6pt minus6pt
    \bigbreak
    \global\advance\subsubno by 1
    \noindent {\sl \the\secno.\the\subno.\the\subsubno. #1}\null. }
%
%
\def\appendix#1
   {\vskip0pt plus.1\vsize\penalty-250
    \vskip0pt plus-.1\vsize\vskip24pt plus12pt minus6pt
    \subno=0
    \global\advance\appno by 1
    \noindent {\bf Appendix \the\appno. #1\par}
    \bigskip
    \noindent}
%
%
\def\subappendix#1
   {\vskip-\lastskip
    \vskip36pt plus12pt minus12pt
    \bigbreak
    \global\advance\subno by 1
    \noindent {\sl \the\appno.\the\subno. #1\par}
    \nobreak
    \medskip
    \noindent}
%
%
\def\ack
   {\vskip-\lastskip
    \vskip36pt plus12pt minus12pt
    \bigbreak
    \noindent{\bf Acknowledgments\par}
    \nobreak
    \bigskip
    \noindent}
%
%

%
%
\def\tabcaption#1
   {\global\advance\tableno by 1
    \noindent {\bf Table \the\tableno.} \rm#1\par
    \bigskip}
%
%
\def\figures
   {\vfill\eject
    \noindent {\bf Figure captions\par}
    \bigskip}
%
%
\def\figcaption#1
   {\global\advance\figureno by 1
    \noindent {\bf Figure \the\figureno.} \rm#1\par
    \bigskip}
%
%
\def\references
     {\vfill\eject
     {\noindent \bf References\par}
      \parindent=0pt
      \bigskip}
%
%
\def\refjl#1#2#3#4
   {\hangindent=16pt
    \hangafter=1
    \rm #1
   {\frenchspacing\sl #2
    \bf #3}
    #4\par}
%
%
\def\refbk#1#2#3
   {\hangindent=16pt
    \hangafter=1
    \rm #1
   {\frenchspacing\sl #2}
    #3\par}
%
%
\def\numrefjl#1#2#3#4#5
   {\parindent=40pt
    \hang
    \noindent
    \rm {\hbox to 30truept{\hss #1\quad}}#2
   {\frenchspacing\sl #3\/
    \bf #4}
    #5\par\parindent=16pt}
%
%
\def\numrefbk#1#2#3#4
   {\parindent=40pt
    \hang
    \noindent
    \rm {\hbox to 30truept{\hss #1\quad}}#2
   {\frenchspacing\sl #3\/}
    #4\par\parindent=16pt}
%
%
\def\dash{---{}---}
%
%
\def\frac#1#2{{#1 \over #2}}
%
%

%
%
\def\d{{\rm d}}
%
%
\def\e{{\rm e}}
%
%
\def\i{\ifmmode{\rm i}\else\char"10\fi}
%
%

%
%

%
%

%
%
\def\etal{{\sl et al\/}\ }
%
%
\catcode`\@=11
%
%
\def\ind{\hbox to 5pc{}}
%
%
\def\eq(#1){\hfill\llap{(#1)}}
%
%

%
%
\def\deqn#1{\displ@y\halign{\hbox to \displaywidth
    {$\@lign\displaystyle##\hfil$}\crcr #1\crcr}}
%
%
\def\indeqn#1{\displ@y\halign{\hbox to \displaywidth
    {$\ind\@lign\displaystyle##\hfil$}\crcr #1\crcr}}
%
%
\def\indalign#1{\displ@y \tabskip=0pt
  \halign to\displaywidth{\ind$\@lign\displaystyle{##}$\tabskip=0pt
    &$\@lign\displaystyle{{}##}$\hfill\tabskip=\centering
    &\llap{$\@lign##$}\tabskip=0pt\crcr
    #1\crcr}}
\catcode`\@=12
%
%



%
%



%
%

%
%
\def\\{\hfill\break}
\def\avd#1{\overline{#1}}
\def\avt#1{\left\langle#1\right\rangle}

\def\cit#1{[#1]}

\def\etal{{\it et al.}}
\def\Ga0{\Gamma_o}
\def\Ja{J^{\rm A}}
\def\Js{J^{\rm S}}

\def\s{\sigma}
\def\sgn{{\rm sign}\,}


\title{Transition from regular to complex behaviour in a discrete
       deterministic asymmetric neural network model}

\author{A Crisanti\dag, M Falcioni\dag\ddag\ and A Vulpiani\dag\ddag}

\address{\dag\ Dipartimento di Fisica, Universit\`a di Roma
                              ``La Sapienza'', I--00185 Roma, Italy}
\address{\ddag\ INFN Sezione di Roma}


\beginabstract
We study the long time behaviour of the transient before the collapse on
the periodic attractors of a discrete deterministic asymmetric neural
networks model. The system has a finite number of possible states so
it is not possible to use the term chaos in the usual sense of sensitive
dependence on the initial condition. Nevertheless, at varying the asymmetry
parameter, $k$, one observes a transition from ordered motion (i.e. short
transients and short periods on the attractors) to a ``complex'' temporal
behaviour. This transition takes place for the same value $k_{\rm c}$ at
which one has a change for the mean transient length from a power law in the
size of the system ($N$) to an exponential law in $N$. The ``complex''
behaviour during the transient shows strong analogies with the chaotic
behaviour: decay of temporal correlations, positive Shannon entropy,
non-constant Renyi entropies of different orders. Moreover the transition
is very similar to that one for the intermittent transition in chaotic
systems: scaling law for the Shannon entropy and strong fluctuations of
the ``effective Shannon entropy'' along the transient, for $k > k_{\rm c}$.
\endabstract

\section{Introduction}
The properties of the low temperature phase of spin systems with random
quenched couplings have been object of intesive analysis in the last years
\cit{M\'ezard \etal 1987, Fischer and Hertz 1991}.
The interest for such models comes from the study of both spin glasses and
neural networks \cit{Amit 1989}. One of the best known among these models,
the Serrington~-~Kirkpatrick (SK) model, is a fully connected version of a
spin glass. Here the interactions between pairs of spins are random, quenched
and symmetric. This leads to a large number of metastable states which are
responsible for the complicated dynamical behaviour of this and related models.

{}From a biological point of view the constraint of symmetric interactions is
unsatisfactory. In fact, in the language of neural networks, two neurons will
not act on each other in a symmetric way. It is then natural to investigate
models with asymmetric couplings between spins \cit{Amit 1989}.

Apart from these biological motivations, the study of disordered spin
systems with asymmetric bonds is also an interesting problem from the
viewpoint of non equilibrium statistical mechanics. In fact for these models
one no longer has a Hamiltonian and a detailed balance condition, and
therefore no fluctuation dissipation theorem exists
\cit{Crisanti and Sompolinsky, 1987}. This last relation
between the response and the correlation functions is, however, essential for
extracting the long time properties of the dynamics. As a consequence, the
long time limit has to be calculated via the full dynamic problem and cannot
be evaluated by equilibrium statistical mechanical averages.

Till now such models have been studied principally in the framework of
neural networks model, and the interest has been mainly focused on the
properties of the attractors of networks of spins with asymmetric couplings.
In particular the effect of the degree of asymmetry on the long time
behaviour has been analysed. Although analytical results, based on
perturbation expansions in the nonlinearity parameter of soft spin models
\cit{Crisanti and Sompolinsky 1987},
showed a drastic change in the deterministic dynamics as soon as the
asymmetry was switched on, subsequent numerical data are in agreement with a
transition at a finite degree of asymmetry \cit{Spitzner and Kinzel 1989}.
A perturbation expansion in the asymmetry parameter for the hard Ising spin
seems to suggest a transition at a finite value of the asymmetry parameter
\cit{Rieger \etal 1989}. The location of the transition is, however,
controversial.

An interesting feature of these models is the presence of extremely long
transients \cit{Crisanti and Sompolinsky 1988, N\"utzel 1991}, i.e. the time
the system wanders around before it settles down in
the ``asymptotic'' regime. It has been found that while for
nearly-symmetric networks the typical number of updates, before the relaxation
onto the attractor, grows as a power of the system size, for strong enough
asymmetry the transients are typically exponential in the size of
the system \cit{N\"utzel 1991}. Therefore it is natural to study the
properties of the  transient, because they give the features of a large
enough  system during any reasonable time.

In this paper we investigate the deterministic parallel dynamics of a
fully connected Ising spin system with random quenched asymmetric
interactions. We stress that if the spins are continuous, then this system
exhibits chaotic behaviour, e.g. positive Lyapunov exponent, in the limit of
high asymmetry \cit{Sompolinsky \etal 1988, Tirozzi and Tsodyks 1991}.
In the case of Ising spins, however, because of the discrete nature of the
states, it is not possible to characterize the time behaviour of the system
by a straigthforword application of the most simple methods of chaotic
dynamics, e.g. by the computation of the Lyapunov exponents.
Nevertheless one can perform a study of the temporal evolution in terms of
well defined quantities, such as Shannon and Renyi entropies, which
give a non ambiguous degree of the ``complexity'' of the system.
{}From these indicators and from the behaviour of the correlation function
one has a strong evidence for a transition, at varying the asymmetry
parameter, from regular to ``complex'' behaviour in systems of finite size.
This transition is very  similar to the onset of turbulence in the
intermittent scenario in chaotic dynamical systems
\cit{Pomeau and Manneville 1980}.

The model will be described in Sect.~2; Sect.~3 deals with
the chaotic --- i.e. complex, in the information theory meaning ---
behaviour during the transient and in Sect.~4 we study the fluctuations of the
chaotic degree along the trajectories, in terms of the Renyi entropies.
In Sect.~5 the reader will find a summary and conclusions.

\section{The model}
In this paper we consider an asymmetric spin glass model with deterministic
dynamics \cit{Crisanti and Sompolinsky 1987}. The model consists of $N$ fully
connected Ising spin $\s=\pm 1$ interacting via quenched random asymmetric
couplings of the form
   $$J_{ij}= \Js_{ij} + k\,\Ja_{ij}, \qquad k\ge 0
    \eqno(2.1)$$
where $\Js_{ji}=\Js_{ij}$ and $\Ja_{ji}=-\Ja_{ij}$, and $(i,j)$ denotes a
pair of spins. The diagonal elements $\Js_{ii}$ are zero.
The elements of the upper triangular part of the symmetric and
antisymmetric matrices $\bss{J}^{\rm S}$ and $\bss{J}^{\rm A}$ are random
independent Gaussian variables with zero mean and mean square equal to
$1/(N-1)(1+k^2)$ so that the rows of the couplings matrices have
average norm $1$ independently of the system size or symmetry.

The parameter $k$ defines the symmetry of the coupling, and ranges
from $k=0$ for symmetric through $k=1$ for asymmetric to $k=\infty$
for antisymmetric couplings. In the literature it is often used the
related parameter $\lambda$ defined by
   $$\lambda= {1-k^2\over 1+k^2}= N\,\left[J_{ij}\,J_{ji}\right]
    \eqno(2.2)$$
ranging from $\lambda=1$ for symmetric through $\lambda=0$ for asymmetric to
$\lambda=-1$ for antisymmetric matrices. In (2.2) square
brakets denote average over the realization of the coupling matrix $\bss{J}$.

Our analysis is restricted to deterministic parallel dynamics, i.e.
the spins are updated simultaneously according to
   $$\s_i(t+1)= \sgn\left[\sum_{j=1}^N\, J_{ij}\,\s_j(t) \right].
    \eqno(2.3)$$
For any given sample $\bss{J}$ the dynamics is deterministic and
depends only on ${\bf S}(0)$ the spin configuration $\{\s_i\}$
at time $t=0$.

For a system of $N$ spins there are $2^N$ different possible states. The
evolution (2.3) connects these states among them. Therefore for each
realization of the couplings the structrure of the dynamics can be depicted
in terms of oriented graphs, i.e. a set of points representing the states
connected by arrows indicating the transitions. An example is shown in
Fig.~1. Obviously each state has one and only one outgoing arrow, but
different arrows can end at the same state. Since the number of states is
finite and the dynamics is deterministic, each initial configuration of
spins evolves to a definite attractor, which can be either a fixed point
${\bf S}(t+1)= {\bf S}(t)$, or a periodic repetition of $l$ configurations
${\bf S}(t+l)= {\bf S}(t)$ (cycle of length $l$).

{}From this remark one may be tempted to conclude that the behaviour of (2.3)
is rather ``trivial'' in the sense that there is no possibility of having
something similar to a chaotic behaviour. This is not completely true. In
fact (2.3) can be seen as a cellular automaton, for which the behaviour just
described is common. Nevertheless, it is well known that there is a wide
class of cellular automata with rather ``complex'' behaviours.

The reason is that the time it takes to reach the attractor can be very long.
As a consequence the system may exhibit for a long time a rather complex
behavior before it eventually relaxes to the ``asymptotic'' regime,
usually a cycle. The presence of long transients, exponential in the system
size, was first observed for sequential dynamics
\cit{Crisanti and Sompolinsky 1988} and recently analysed in more details for
both sequential and parallel dynamics \cit{N\"utzel 1991}.

{}From the above considerations it follows that in these models there are two
relevant time scales: the relaxation time $\tau$ and the cycle length $l$.
These are defined as
   $$l= \min_{n}\,\bigl[\,{\bf S}(t+n)= {\bf S}(t)\,\bigl]
    \eqno(2.4)$$
   $$\tau= \min_{n}\,\bigl[\,{\bf S}(l+n)= {\bf S}(n)\,\bigl].
    \eqno(2.5)$$
The time $\tau$ is the time the system needs to reach the first cycle of
length $l$. In general both $l$ and $\tau$ depend on the initial state
${\bf S}(0)$ and on the coupling realization $J_{ij}$. For any given
$k$ and system size N one can define the average transient time and the
average cycle length. To this end one considers a large number $M$ of
independent coupling realizations $J_{ij}$. For each sample and a random
initial state ${\bf S}(0)$ the time $\tau$ and the cycle length $l$ are
recorded. One then defines the average values as
   $$\avt{l}= {1\over M}\, \sum_{m=1}^M\, l(m)
    \eqno(2.6)$$
   $$\avt{\tau}= {1\over M}\, \sum_{m=1}^M\, \tau(m)
    \eqno(2.7)$$
where $l(m)$ and $\tau(m)$ are the values of $l$ and $\tau$ defined in (2.4)
and (2.5) for the $m$-th sample.

Studying these average quantities as a function of both the symmetry and the
system size, N\"utzel found strong numerical evidence for a transition
between two different behaviours as the symmetry of the couplings is varied.
He found that for high symmetric couplings $\avt{l}$ is size independent
and $\avt{\tau}$ grows as a power of $N$, while the behaviour of both
quantities is exponential in $N$ for high asymmetric couplings.
He located the transition point around $\lambda\simeq 0.5$ where $\avt{l}$
shows strong changes \cit{N\"utzel 1991}. By contrast $\avt{\tau}$ changes
smoothly with $\lambda$. This behaviour was found for both sequential and
parallel dynamics.

The values of $l$ and $\tau$, however, exhibit strong sample to sample
fluctuations. As a consequence the leading contributions to the average
(2.6) and (2.7) come from rare events. This suggests that the
appropriate quantity for averaging is not $\tau$ and $l$, but their
logarithms. One then defines  the ``typical'' values
   $$\tau_{\rm typ}= \exp\avt{\ln\tau}
    \eqno(2.8)$$
   $$l_{\rm typ}= \exp\avt{\ln l}.
    \eqno(2.9)$$
An analysis of these quantities as a function of both the symmetry of the
couplings and the system size reveals that
   $$\tau_{\rm typ}\sim\avt{\tau}\propto
         N^{\alpha(k)}\,\e^{\beta(k)\,N}
    \eqno(2.10)$$
where $\beta(k)\ll 1$ for $k_{\rm c}\approx 0.5$. A similar behaviour
is observed for the cycle length $l$.

We find that, for $k \geq k_{\rm c}$, $\tau$ has a roughly lognormal
distribution whose peak and width scale linearly with $N$. In other
words $\ln\tau$ is a gaussian with
   $$\avt{\ln\tau}\propto N
    \eqno(2.11a)$$
   $$\avt{(\ln\tau)^2} - \avt{\ln\tau}^2\propto N^2.
    \eqno(2.11b)$$
The same result remains true also for sequential dynamics
\cit{Crisanti and Sompolinsky 1988}.  The lognormal distribution is quite
common in disordered systems \cit{Paladin and Vulpiani 1987}. However, in
these models one has a non standard scaling of the
variance of $\ln\tau$. In fact usually both the mean value and the variance
scale with $N$. This means that here there are enormous fluctuations:
the ratio between the standard deviation and the mean value is constant
instead of being proportional to $N^{-1/2}$ as in the standard case.

The presence of these long transients, e.g. for $k=0.8$ and
$N=100$ one has $\tau_{\rm typ}= O(10^6)$, makes natural to study the
dynamics during the transient. Not only because this is the behaviour
of the system on reasonable times, but also because this is the behaviour
which is revealed by any mean-field study of these systems.

To analyze the temporal evolution one of the first quantities one usually
studies are the time correlations. For a given initial state ${\bf S}(0)$
and coupling realization $J_{ij}$ we compute the following quantity
   $$C(t)= {1\over N}\, \sum_{i=1}^N\,\avd{\s_i(t+t')\,\s_i(t')}
    \eqno(2.12)$$
where the bar denotes time average. The correlation $C(t)$ is evaluated
taking into account only the data for the transient, i.e. times less than
$\tau$. We remark that $C(t)$ is not averaged over different
realizations of $J_{ij}$. From the comparison of several realizations
and different $N$, it seems that $C(t)$ is a selfaveraging quantity
that depends very weakly on $N$. On the contrary
$c_i(t)= \avd{\s_i(t+t')\,\s_i(t')}$ changes at varying $i$; and there
exist two different possibilities: $c_i(t)$ either decreases
monotonically with $t$, or it shows strong oscillations with a decreasing
envelope.

For $k=1$ (full asymmetric case) one has $C(t)=0$ for $t\geq 1$.
The behaviours of $C(t)$ for $k<1$ are reported in Fig. 2a.
Unfortunatelly one does not have a simple formula to fit $C(t)$ as a function
of $t$ at different $k$, even if, for each $k$, the data seem consistent
with a stretched exponential decay:
   $$C(t)\sim \exp[ -a\,t^b] \,\,\,\, {\rm with} \,\,\,\, b<1 .
    \eqno(2.13)$$
Nevertheless one can define a characteristic time, $\tau_{\rm c}$, as
the time at which the envelope of $C(t)$ reaches a threshold value,
e.g. $1/\e$. Figure 2b shows $1/\tau_{\rm c}$ vs $k$; one has evidence
for a transition at $k\approx 0.5$. The above scenario does not
change using different values for the threshold.

\section{Chaotic behaviour during the transient}

It is known \cit{Isola \etal 1985} that in some dynamical systems with many
degrees of freedom the transition from regular to chaotic motion is also
signaled by a change, from power to exponential in the number of degrees of
freedom, of the Poincar\`e return time.

This result and the analysis of the correlation functions discussed in the
previous section suggest the possibility of a ``chaotic behaviour''
during the transients for enough asymmetric couplings. This is in spite of
the fact that a stationary periodic orbit is always found to be the final
attractor. The reason to study in more details this chaotic behaviour follows
from the fact that if the asymmetry is strong enough the transient increases
exponentially with the system size. Therefore any practical investigation of
the time evolution is necessarily limited to the transient. Moreover one can
argue that the attractor might not be relevant for the dynamics since even in
the presence of stable solutions, generic initial conditions relax towards it
only after exponentially long transients \cit{Crutchfield and Kaneko 1988}.

Due to the discrete nature of the spin and to the full connectivity of the
system a quantification of ``chaos'' during the transient is not at all
straightfowrard. In fact, for systems whose state changes continuously one
can quantify the degree of chaos in terms of the Lyapunov exponents
\cit{Benettin \etal 1980a, b}. They measure the exponential growth rate of
the distance between two initially close trajectories.

One can try to follow the same idea of the Lyapunov exponents
studing two trajectories whose initial condition ${\bf S}(0)$ and
${\bf S}'(0)$ differ by a single spin.
Looking at the growth of the Hamming distance between ${\bf S}(t)$ and
${\bf S}'(t)$ one can then introduce something similar to the first Lyapunov
exponent.
This is the socalled `damage spreading' method which revealed useful for
discrete systems similar to (2.3) but with only local interactions among the
spins \cit{Kauffman 1969, Derrida and Stauffer 1986}.

In this paper to characterize the chaotic behaviour we use a completely
different
approach. Following the basic idea of Shannon in the information theory
\cit{Khinchin 1957}, we look at the ``complexity'' of the temporal history on
the $i$-th spin during the transient \cit{Grassberger 1986}. For each initial
configuration and fixed realization of $J_{ij}$ the ``history life'' of a
spin $\s_i$ before the system relaxes to the attractor can be written as a
sequence
   $$\s_i(0),\,\s_i(1),\,\ldots,\,\s_i(m)
    \eqno(3.1)$$
where $m< \tau + l$. Since the transient increases exponentially with the
system size, in principle the length of the sequence (3.1) can be done as
large as we want by increasing $N$.

The complexity of the temporal behaviour of the spin can be caracterized by
looking at the average amount of information contained in the sequence (3.1).
To this end we consider a sequence $\omega_n$ of $n$ possible outcomes of the
spin $\s_i$. Since it can assume only two values there are $2^n$ possible
sequences. These can be regarded as strings of $n$ binary digits.
In general, the sequences $\omega_n$ will appear in the history life
of the spin $\s_i$ with different frequencies. Some of them may appear not
at all or with very small probability. By computing the frequency with
which each $\omega_n$
appears in the sequence (3.1) we can define the probability $P(\omega_n)$
that the dynamics generates the sequence $\omega_n$. The measure of
information contained in the knowledge of the $P(\omega_n)$ is given by the
entropy $H(n)$ defined as
   $$H(n)= -\sum_{\omega_n}\, P(\omega_n)\, \ln P(\omega_n)
    \eqno(3.2)$$
where the sum is over all the sequences $\omega_n$. If the sequence (3.1) is
stationary the quantity $H(n)$ depends only on $n$ and not on the initial
state $\s_i(0)$. In this case, in the limit of large $n$ the average
information $H(n)/n$ converges to the limit value $h$ which is called the
Shannon entropy of the sequence (3.1) \cit{Khinchin 1957}. The existence of
this limit implies that for large $n$ the mean gain of information in
considering sequences of length $n+1$ instead of $n$, i.e.
   $$h(n)= H(n+1) - H(n)
    \eqno(3.3)$$
approaces a constant limit given by the Shannon entropy $h$.

If the sequence (3.1) is a Markov chain, then $h(n)= h$ for each $n$. This
can be generalized to Markov chains of order $p$, i.e. the probability of
having an event at time t depends only on the events at times $t-p$ up to
$t-1$, in which case $h(n)=h$ for $n\ge p$.

The relevance of $h$ follows from the first Shannon~--~McMillan theorem
\cit{Khinchin 1957} which states that for large $n$
the number of $n$-term sequences $\omega_n$ one can really observe in
(3.1) is
   $$N_{\rm eff}(n) \sim \e^{h\, n}.
    \eqno(3.4)$$
This can also be stated as follows. For $n$ sufficiently large all the
$n$-term sequences $\omega_n$ can be separated into two classes $\Omega_1(n)$
and $\Omega_2(n)$, such that for every sequence in the first class
$P(\omega_n)\simeq \exp(-hn)$ and
   $$\sum_{\omega_n \in\Omega_1(n)} P(\omega_n) \to 1
       \quad \hbox{\rm for} \quad  n\to \infty.
    $$
On the contrary, for the sequences in the second class
   $$\sum_{\omega_n \in\Omega_2(n)} P(\omega_n) \to 0
     \quad \hbox{\rm for} \quad   n\to \infty.
    $$

If we remember that the maximum number of $n$-term sequences is
   $$N_{\rm max}(n)=2^n=\e^{n\,\ln 2}
   $$
we see that if $h< \ln 2$ for large $n$ the class $\Omega_1(n)$ contains only
a negligible fraction of sequences. On the contrary, the overwhelming majority
of such sequences fall into $\Omega_2(n)$.

If the probability of having $\s_i=1$ or $\s_i=-1$ in (3.1) is the same, as
it is in our case, then $h<\ln 2$ signals the presence of a rule in the
sequence. Thus the Shannon entropy gives a measure of the ``complexity'' of
the sequence. In particular if $h=0$ then we are in presence of periodic
motion, while if $h= \ln 2$ the sequence is generated with an head and tail
trial. We stress that $\ln 2$, the largest value attainable by $h$, is
reached only if the symbols in (3.1) are independent and equiprobable. In
fact $h< \ln 2$ also if the symbols are independent but not equiprobable,
e.g. generated according an head and tail trial done with an unbalanced coin
\cit{Khinchin 1957}.

To analyze the motion of the system during the transient, we have computed
the Shannon entropy of different spins for a fixed, randomly chosen,
configuration of couplings $J_{ij}$. In all the cases we have taken
$N$ large enough to ensure sufficiently long transients:
$\tau\sim O(10^5)$ at least. For for $k\ge 0.7$ we considered $N=100$,
$N=200$ and $N=300$, while for $k<0.7$ we used $N=200$ and $N=300$.
For smaller value of $k$ the required system size becomes very large
and, since the computational time increases as $N^2$, it is
practically impossible to reach too small values of $k$. In all the cases we
checked that the sequence (3.1) for each spin is stationary.

By computing $h(n)$ for different $n$ we found a rapid convergence towards the
Shannon entropy $h$. In Fig.~3 it is shown a typical case from which we see
that $n=4$ or $n=5$ are sufficient to have a good extimate of $h$.

In the limit case of $k=1$ the evaluated $h$ is very close to the largest
value $\ln 2$. This is not surprising since if the couplings $J_{ij}$ and
$J_{ji}$ are uncorrelated then the statistical properties of the dynamical
system (2.3) are the same as the sequence generated by an head and tail
trial \cit{Gutfreund \etal 1988}.

For a given configuration of couplings $J_{ij}$ and a fixed value of $N$
we have computed the Shannon entropy $h^{(i)}$ for each sequence (3.1). The
mean value
     $$\avt{h} ={1\over N}\, \sum_{i=1}^N\, h^{(i)} $$
seems to be a selfaveraging quantity, which does not depend on $N$;
moreover the variance
     $${1\over N}\, \sum_{i=1}^N\, \left( h^{(i)} - \avt{h} \right) ^2
      $$
decreases for increasing $N$. In Fig. 4 we show $\avt{h}^2$ as a function of
$k$: it is  well evident the scaling
     $$\avt{h}\sim \left( k-k_{\rm c}\right) ^{{1\over 2}}
        \,\,\,\, {\rm with} \,\,\,\,
       k_{\rm c} = 0.50 \eqno(3.5)$$

We close this section with a remark. Note that the point at $k=0.56$
does not agree very well with Eq.~(3.5). Since it is practically
impossible for us to work with smaller values of $k$, we cannot say
if this point can be considered as an evidence for a crossover
behaviour. In Sect.~5 we shall consider again this issue.

\section{Fluctuations during the transient}

In general not all the $2^n$ possible $n$-term sequences will appear in the
sequence (3.1) generated by the dynamics. Nevertheless, for large $n$ their
number $\widetilde{N}(n)$ increases exponentially with $n$. The rate of
growth
   $$K_0= \lim_{n\to \infty} {1\over n} \ln \widetilde{N}(n)
    \eqno(4.1)$$
is the topological entropy. The smaller $K_0$ the larger the number
of forbidden $n$-term sequences in (3.1). For our system we found that at any
value of $k$ the topological entropy assumes its maximum value
$K_0=\ln 2$. This means that for fixed $J_{ij}$ and generic initial
condition, all possible sequences are actually generated by (2.3).

The topological entropy and the Shannon entropy are global quantities, since
they ignore the finite time, i.e. finite $n$,  fluctuations.  For any
finite $n$ we can define the ``effective Sahnnon entropy'' $\gamma(n)$ of the
$n$-term string $\omega_n$ as
  $$\gamma(n)=- {1 \over n} \ln P(\omega_n).
   \eqno(4.2)$$
In general the value of $\gamma(n)$ will depend on the string $\omega_n$, even
for large but finite $n$. These finite $n$ fluctuations can be characterized
by means of the Renyi entropy \cit{Renyi 1970, Paladin and Vulpiani 1987}.
Let us define
   $$H_{q}(n)= -\,{1\over q-1}\, \ln\left[\sum_{\omega_n}P(\omega_n)^q
                                    \right]
    \eqno(4.3)$$
where the sum is extended over all the $2^n$ $n$-term sequences $\omega_n$.
In the limit of large $n$ the quantity
   $$h_{q}(n)=H_{q}(n+1)-H_{q}(n)
    \eqno(4.4)$$
or equivalently $H_q(n)/n$, converges towards the limit $h_q$ called
Renyi entropy of order $q$.

{}From the definition it readly follows that Shannon and topological entropies
are obtained from $h_q$ via $h= \lim_{q \to 1}h_q$ and
$K_0= \lim_{q \to 0}h_q$. It is possible to show that in general $h_q$ is a
not-increasing function of $q$ \cit{Renyi 1970}. In the limit case of absence
of fluctuations one has $h_q=K_0= \hbox{\rm constant}$. Thus the deviation of
$h_q$ from  $K_0$ gives a measure of the relevance of the fluctuations.

As we have done in Sect.~3 for the Shannon entropy, we have computed,
for each sequence $\s_i(1),\,\s_i(2),\,\ldots$, the Renyi entropies
$h_q^{(i)}$. For values of $k$ close to $1$ $h_q^{(i)}$ does not change
very much at varying $q$, while for $k$ close to $k_{\rm c}$ one observes
large deviations of $h_q^{(i)}$ from $h^{(i)}$. In Fig. 5 we show
$\avt{h_q}$ as a function of $q$ for $k=0.9$ and $k=0.7$: even at a
qualitative level, the difference is well evident.
The properties discussed in Sect.~3,
for the Shannon entropy, hold true also for $h_q$: each $\avt{h_q}$ is a
selfaveraging quantity which does not depend on $N$.

The Renyi entropies are deeply related to the probability distribution
$P_n(\gamma)$ of the ``effective Shannon entropy'' $\gamma(n)$.
For large $n$ it is reasonable to make the ansatz that $P_n(\gamma)$
vanishes exponentially \cit{Eckmann and Procaccia 1986, Paladin \etal 1986},
   $$P_n(\gamma) \propto \exp\bigl[-n\, S(\gamma)\bigr],
    \qquad S(\gamma)\ge 0.
    \eqno(4.5)$$
The equal sign holds only for $\gamma=h$, as a consequence of the
Shannon-McMillan theorem. With this assumption one finds
   $$h_q= {1 \over q-1}\,
            \min_{\gamma}\bigl[\, (q-1)\, \gamma + S(\gamma)\,\bigl].
    \eqno(4.6)$$
This is a Legendre transform which shows that each $q$ selects a
particular class of sequences with ``effective Shannon entropy'' $\gamma$.
In particular $q=1$ selects the most probable sequences, whereas
for $q\not=1$ the leading contribution to $h_q$ comes from rare
sequences.

For $\gamma \simeq h$ one has the following parabolic approximation for
$S(\gamma)$
   $$S(\gamma)={(\gamma -h)^2 \over 2 \mu},
    \eqno(4.7)$$
where $\mu$ is given by
$$\eqalign{\mu=& \lim_{n\to \infty} {1\over n}\left[ \sum_{\omega_n}\,
            P(\omega_n)\, \big[\ln P(\omega_n)\big]^2 -
            \big[\sum_{\omega_n}\, P(\omega_n)\,
             \ln P(\omega_n)\big]^2 \right]  \cr
    =&\lim_{n\to \infty} n \int (\gamma-h)^2 P_n(\gamma)\d\mu .\cr} $$
Equation (4.7) corresponds to
   $$h_{q}=h- \mu\,(q-1) / 2, \qquad q\simeq 1;
    \eqno(4.8)$$
one can consider $\mu$ as an inhomogeneity factor.
The (4.7) and (4.8) correspond to a gaussian approximation. The quantities
$h$ and $\mu$ are the most relevant caracterizations for the "complexity"
of the sequence since $h$ gives the typical value and $\mu$ is related
to the variance of the fluctuations.

We found that for $k$ close to $1$ $\mu$ assumes very small values,
indicating that the sequences are almost homogeneous. When $k$ is
decreased $\mu$ increases and assumes its largest values
(strong fluctuations) as one reaches the critical value $k_{\rm c}$
In Fig.~6 we show $\avt{\mu}$ as a function of $k$.

It is interesting to note that the behavior of $\avt{h}$ and $\avt{\mu}$
as function of the parameter $k$ is similar to the one observed for the
intermittent transition in dymical system, e.g. in the Lorenz model
\cit{Pomeau and Manneville 1980}.

We close this section noting that all the results for $\avt{h}$ and
$\avt{\mu}$ for $k$ larger than approximately $0.5$ are practically
independent of $N$ in the range $N=50\div 500$.

\section{Summary and conclusions}
In this paper we have studied the long-time behaviour of binary
sequences, obtained from the deterministic evolution of an asymmetric
neural networks model with $N$ components. The system size $N$ is large but
finite. In spite of the discrete nature of its states, at varying the
asymmetry parameter $k$, one observes some features very similar
to the intermittent transition in chaotic systems. For $k<k_{\rm c}$
one has that the typical lenght of the transient follows a power law
in the size of the system $N$, while for $k>k_{\rm c}$ the transients
increase exponentially with $N$. One can define, in an unambiguous way,
some degree of ``complexity'' during the transient, by using well
established concept (Shannon and Renyi entropies) in the information
theory.

For $k>k_{\rm c}$ the time behaviour of the transient is very similar
to that one in the chaotic systems: decay of the temporal correlation
and positive Shannon entropy. Around $k_{\rm c}$ one has a scaling law
for the mean Shannon entropy $\avt{h}$:
$$\avt{h} \sim \sqrt{k-k_{\rm c}},$$
moreover the fluctuations of the degree of chaos, given by
the mean inhomogeneity factor $\avt{\mu}$, are large for $k$
approaching $k_{\rm c}^+$.

All our results on the Shannon and Renyi entropies and characteristic times
for $k$ larger than approximately $0.5$ do not depend on the value of $N$ in
the investigated range $N=50\div 500$. However, in
principle, we cannot exclude the possibility that what we found is
not a real sharp transition, but the manifestation, at finite $N$,
of a crossover behaviour. For instance, for the sequential
dynamics \cit{Crisanti and Sompolinsky 1988}
one has an exponential growth with $N$ for $\tau_{\rm typ}$
according to (2.10) even for $k<k_{\rm c}$,
but the constant $\beta (k)$ is so small that in order to see the
exponential law one has to consider enormous systems, $N=O(10^3)$.
In a symilar way in the above quoted reference the average correlation
function during the transient has a characteristic time $\tau_{\rm c}$
proportional to $k^{-6}$ for small $k$.

{}From our computations and the previous study of N\"utzel \cit{1991} one can
conclude that the dependence of $k_{\rm c}$ on $N$, if any, is very weak.
For almost all the practical purposes, however, the nature of the transition,
i.e. sharp or crossover, is not very important. Nevertheless, its
existence is relevant for numerical studies. For example, it could be very
dangerous to compute temporal averages at different $k$ using the same $N$
and maximal time. Indeed the meaning of these averages is completely
different: for $k> k_{\rm c}$ one should have informations about the
transient while for $k<k_{\rm c}$ one just observes properties of the
periodic attractors.

We conclude by noting that similar problems are
present also in hamiltonian (and more generally symplectic) systems
with many degrees of freedom \cit{Falcioni \etal 1991}.

\ack
We thank D J Amit, F Bagnioli, H Kunz and S Ruffo for useful discussions and
suggestions.

\references

\refbk{Amit D J 1989}{Modeling brain function}%
      {(Cambridge University Press)}

\refjl{Benettin G, Galgani L, Giorgilli A and Strelcyn J M 1980a}%
      {Meccanica}{15}{9}

\refjl{\dash 1980b}{Meccanica}{15}{21}

\refjl{Crisanti A and Sompolinsky H 1987}{Phys Rev A}{36}{4922}

\refjl{\dash 1988}{Phys Rev A}{37}{4865}

\refjl{Crutchfield J P and Kaneko K 1988}{Phys Rev Lett}{60}{2715}

\refjl{Derrida B and Stauffer D 1986}{Europhys Lett}{2}{739}

\refjl{Eckmann J-P and Procaccia I 1986}{Phys Rev A}{34}{659}

\refjl{Falcioni M, Marini Bettolo Marconi U and Vulpiani A 1991}%
      {Phys Rev A}{44}{2263}

\refbk{Fischer K H and Hertz J A 1991}{Spin Glasses}%
      {(Cambridge University Press)}

\refjl{Grassberger P 1986}{Int J Theor Phys}{25}{907}

\refjl{Gutfreund H, Reger J D and Young A P 1988}{J Phys A}{21}{2775}

\refjl{Isola S, Livi R and Ruffo S 1985}{Phys Lett A}{112}{448}

\refjl{Kauffman S A 1969}{J. Theor. Biol.}{22}{437}

\refbk{Khinchin A Ya 1957}{Mathematical Foundations of Information Theory}%
      {(Dover - New York)}

\refbk{M\'ezard M, Parisi G and Virasoro M A 1987}%
             {Spin Glass Theory and Beyond}{(World Scientific)}

\refjl{N\"utzel K 1991}{J Phys. A}{24}{L151}

\refjl{Paladin G, Peliti L and Vulpiani A 1986}{J Phys A}{19}{L991}

\refjl{Paladin G and Vulpiani A 1987}{Phys Rep}{156}{147}

\refjl{Pomeau Y and Manneville P 1980}{Commun Math Phys}{74}{189}

\refbk{Renyi A 1970}{Probability Theory}{(North-Holland)}

\refjl{Rieger H, Schreckenberg M and Zittarz J 1989}{Z Phys B}{74}{527}

\refjl{Sompolinsky H, Crisanti A and Sommers H-J 1988}{Phys Rev Lett}%
      {61}{259}

\refjl{Spitzner P and Kinzel W 1989}{Z Phys B}{77}{511}

\refjl{Tirozzi B and Tsodyks M 1991}{Europhys Lett}{14}{727}

\figures
\figcaption{Some examples of possible behaviour. One recognizes two fixed
            points and three cycles of period $2$, $3$ and $4$, respectively.
           }
\figcaption{a) $C(t)$ as a function of $t$ for $k=0.6$ (diamonds) and
            $k=0.8$ (plus); in order to evidenciate the main feature we plot
            $C(t)$ only at even values of $t$.
            b)  $1/\tau_{\rm c}$ as a function of $k$.
            The system size is $N=200$.
           }
\figcaption{$h(n)$ as a function of $n$ for one spin sequence at
            $k=0.7$ and $N=200$.
           }
\figcaption{$\avt{h}^2$ as a function of $k$ for $N=200$;
            the dashed line indicates
            the law $\avt{h} \propto (k-k_{\rm c})^{1/2}$ with $k_{\rm
            c}=0.50$. The horizontal line corresponds to $\avt{h}= \ln 2$.
           }
\figcaption{$\avt{h_q}$ as a function of $q$ for $k=0.9$ (plus) and $k=0.7$
            (cross). The system size is $N=200$.
           }
\figcaption{$\avt{\mu}$ as a function of $k$ for $N=200$.
           }

\bye